# Direct imaging of band profile in single layer MoS$_2$ on graphite: quasiparticle energy gap, metallic edge states and edge band bending

*Chendong Zhang,[1] Amber Johnson,[1] Chang-Lung Hsu,[2] Lain-Jong Li,[2] and Chih-Kang Shih[1,*]*

[1] Department of Physics, University of Texas at Austin, Austin, TX 78712, USA

[2] Institute of Atomic and Molecular Sciences, Academia Sinica, Taipei, 10617, Taiwan

* Correspondence should be addressed to Chih-Kang Shih: shih@physics.utexas.edu

**Abstract:**

Using Scanning Tunneling Microscopy and Spectroscopy, we probe the electronic structures of single layer MoS$_2$ on graphite. We show that the quasiparticle energy gap of single layer MoS$_2$ is 2.15 ± 0.06 eV at 77 K. Combining this with temperature dependent photoluminescence studies, we deduce an exciton binding energy of 0.22 ± 0.1 eV, a value that is lower than current theoretical predictions. Consistent with theoretical predictions, we directly observe the metallic edge states of single layer MoS$_2$. In the bulk region of MoS$_2$, the Fermi level is located at 1.8 eV above the valence band maximum, possibly due to the formation of a graphite/MoS$_2$ heterojunction. At the edge, however, we observe an upward band bending of 0.6 eV within a short depletion length of about 5 nm, analogous to the phenomena of Fermi level pinning of a 3D semiconductor by metallic surface states.

Transition metal dichalcogenides (TMDs) are members of the layered materials like graphite. Unlike graphene, the existence of an energy band gap makes TMDs very attractive as 2D electronic materials. This is exemplified by the successful demonstration of single layer (SL) $MoS_2$ transistors.[1] The recent demonstration of indirect-to-direct band gap transition of TMDs at single layer thickness, accompanied by a two orders of magnitude enhancement in optical transition efficiency, further inspired significant interest in SL-TMDs as a platform for atomic layer optoelectronics.[2-4] In addition, the valley degeneracy is lifted in SL-TMDs, leading to the interesting valleytronics phenomena.[5,6] Thus far, investigations have utilized primarily devices for transport measurements, optical measurements which have revealed the exciton and trion transitions, and ARPES (angle-resolved photoemission spectroscopy) which has revealed the quasi-particle band structure for *filled electronic states*.[6,7] One important issue that remains unresolved is the magnitude of the exciton binding energy. Optical measurements discovered the trion binding energy of 18 meV (relative to exciton) for $MoS_2$[8] and 30 meV for $MoSe_2$[9] at low temperature, suggesting a large exciton binding energy. First principle calculations predicted an extraordinarily large exciton binding energy, between 0.5 to 1.0 eV.[10-13] Nevertheless, there has been no direct experimental measurement.

Another important issue is the influence of the edge states on the device properties. Early studies of $MoS_2$ triangular nanoclusters (1 – 2 nm in size) on Au substrates showed metallic edge states.[14,15] Larger SL-TMDs 2D islands (micron size) also come in a triangular shape. TEM investigations of $MoS_2$ showed that the edge can either be Mo-terminated, leading to a very straight edge, or S-terminated with a less straight edge.[16,17] More recent calculations showed that both Mo-terminated and S-terminated edges result in metallic states inside the bulk band gap.[16,18,19] Some interesting optical properties have been recently seen along the edge although its origin remains unclear.[17, 20] More importantly, the role of edge states in the understanding of the device characteristics has not yet been investigated.

In order to address these critical issues, we use Scanning Tunneling Microscopy and Spectroscopy (STM/S) to probe the quasiparticle band structure of single layer MoS$_2$ on graphite. We show that in the region away from the edge (i.e. bulk), the MoS$_2$ band profile is homogeneous with a band gap of 2.15 ± 0.06 eV at 77 K. In addition, the Fermi level (E$_F$) is located at 1.83 ± 0.03 eV above the valence band maximum (VBM). Combining this with optical measurements of exciton transitions in the same sample (1.93 eV at 79 K), we deduced an exciton binding energy of 0.22 ± 0.1 eV, a value that is significantly lower than most theoretical calculations.[10-13] Near the edge, the band profile is bent upward by about 0.6 eV. The band bending region is only 5 nm from the edge. At the edge, the metallic nature is also observed from finite conductivity in the gap region. This is a 2D analogy of how metallic surface states pin the Fermi level of a 3D semiconductor.

The SL-MoS$_2$ flakes were synthesized by a vapor phase reaction using MoO$_3$ and sulfur powder as the precursors.[21] In brief, the MoO$_3$ powder (0.6 g) was placed in a ceramic boat located in the heating zone center of the furnace. The sulfur powder was placed in a separate ceramic boat at the upper stream side of the furnace which was maintained at 170 ℃ during the reaction. Highly-oriented-pyrolitic-graphite (HOPG) substrates were placed at the downstream side, while the S and MoO$_3$ vapor were brought to the targeting substrate by a flowing Ar gas (Ar = 70 sccm, chamber pressure = 30 Torr) through the furnace. The center heating zone was heated to 650°C at a ramping rate 25°C/min. After being kept at 650 °C for 10 minutes, the furnace was then naturally cooled down to room temperature. The AFM image (Figure 1(a)) shows that the as-grown sample contains primarily SL-MoS$_2$ flakes with triangular shape formed on graphite surface.

The sample was then transferred into an ultra-high-vacuum (UHV) chamber (base pressure < 6 x 10$^{-11}$ torr), housing a home-made low temperature scanning tunneling microscope. All STM investigations reported here were acquired at 77 K. Electrochemically etched W-tips are cleaned in-situ with electron beam bombardment. Before STM investigations, the samples were cleaned in-situ by heating it up to 250 °C for an extended time (typically longer than 2 hours). Shown in Figure 1(b) is the STM image of a partially cleaned sample (with a lower temperature of 120 °C) where one can see a SL-MoS$_2$ flake that is still partially covered by unknown adsorbates. Similarly, it can be seen that the nearby graphite surface is not clean as well. Shown in Figure 1 (c) is a sample after 10 hours annealing at about 250 °C. The inset line profile indicates that the height of the triangular island is about 0.67 nm, which is consistent with the thickness of SL-MoS$_2$. The close-up image (Figure 1 (d)) shows that the bulk of the SL-MoS$_2$ is free of adsorbates. However, at the edge, small clusters of adsorbates can be found with spacings between clusters ranging from 5 to 20 nm. Even with extended or higher temperature annealing, the edges always contain some small clusters of adsorbates and they appear to be chemically stable. In earlier studies of MoS$_2$ nanoclusters grown on Au(111) with an edge length of few nanometers, high chemical activities were found at the edge.[15,22,23] Thus, we suggest that the adsorbed clusters which we observe on the edge are also due to high chemical activity of the edge, albeit the lateral size is orders of magnitude larger.

The quasi-particle band gap of SL-MoS$_2$ is directly probed using scanning tunneling spectroscopy. Shown in Figure 2(a) is a typical *I-V* curve (shown in black) and *dI/dV-V* curve (shown in red) acquired in the bulk area (far away from the edge) of SL-MoS$_2$. The *dI/dV* spectrum is displayed in logarithmic scale where we marked the apparent location of the valence band maximum (VBM) at a sample bias of -1.81 V and the conduction band minimum (CBM) at +0.30 V (marked by two green dashed arrows). We note however, above this apparent CBM location, there is another threshold located at around 0.5 eV above $E_F$ (marked by the blue dashed arrow). Taking the apparent positions of CBM and VBM, one can get a quasiparticle gap ($E_g$) of 2.11 eV based on this spectrum. As discussed extensively in the past, measuring quasiparticle gap of a large gap semiconductor such as GaAs requires special care.[24] Due to a strong bias dependent transmission probability the "apparent gap" observed in tunneling spectroscopy is often influenced by the junction stabilization bias. Often an "apparent gap" larger than the true gap is observed unless the stabilization voltage is sufficiently close to the location of the band edge, which leads to a smaller tip-to-sample distance.[24] In Fig. 2(a), the stabilization voltage is -2.2 V which is sufficiently close to the VBM (at about -1.8 V). Thus it allows us to detect the band edge accurately.

In Figure 2(b) we show the statistical distributions of the results from more than 80 individual tunneling spectra measured in the bulk area of different monolayer thick flakes. Statistically, the location of VBM has a narrow distribution of 1.83 ± 0.03 eV below $E_F$ while the location of the CBM shows a slightly larger distribution of 0.31 ± 0.05 eV above $E_F$. Also shown in Figure 2(c) is the statistical distribution of the measured quasiparticle band gap, 2.15 ± 0.06 eV, for the single layer $MoS_2$.

To determine the exciton binding energy, we carried out temperature dependent photoluminescence (PL) measurements of the same sample after the STM investigations, and the results are shown in Figure 3. We observed a 1.86 eV optical gap at room temperature which increased to 1.93 eV as we decreased the sample's temperature to 79 K. The PL measurements shown here are not from a single flake, but rather it is an ensemble average of many single monolayer thick MoS$_2$ flakes. Our result is consistent with what has been reported in the literature for SL-MoS$_2$ (1.86 eV optical gap at room temperature).[3] It is also seen that the PL efficiency for SL-MoS$_2$ on graphite is significantly reduced by about two orders of magnitude in comparison to the case of SL-MoS$_2$ on Al$_2$O$_3$. Nevertheless, the optical transition energy remains the same. Comparing the value of 2.15 ± 0.06 eV for quasiparticle gap at 77 K and the value of 1.93 eV for optical gap at 79 K, we can conclude that the exciton binding energy is 0.22 ± 0.1 eV, a value smaller than what has been predicted by theoretical calculations which range from 0.5 to 1.0 eV.[10-13]

As indicated in Fig 2(a), in addition to the apparent CBM position at 0.3 eV above E$_F$, another threshold occurs around 0.5 eV above E$_F$ (marked as a blue dashed arrow and labeled as E$_{th}$). Most (but not all) of the spectra show this second threshold around 0.5 eV. There is a possibility that E$_{th}$ is the actual CBM position while the states below this threshold correspond to an impurity band due to degenerate n-type doping. Nevertheless, the degenerate n-doping should result in a Fermi level above the band edge which is not observed here. On the other hand, it might be possible that the formation of a vertical heterojunction between SL-MoS$_2$ and the graphite (semimetal) pins the Fermi level below the impurity band edge. If this interpretation holds, the position of the CBM would be 0.51 ± 0.02 eV which would lead to a quasiparticle band gap of 2.34 ± 0.05 eV. Then, the corresponding binding energy for exciton is 0.42 ± 0.1 eV, which is closer to the theoretically predicted values. Moreover, as exciton binding energy is likely influenced by the dielectric environment, it will be interesting to investigate how this quantity varies as a function of the supporting substrate.

The spatial dependence of the electronic structure of the SL-MoS$_2$ from the bulk to the edge is mapped out using scanning tunneling spectroscopy. Figure 4(b) shows the path along which 36 tunneling spectra were acquired; the path is represented by a dashed line which is 90 nm long, with spectra taken at 2.5 nm intervals. A subset of the spectra is displayed in Figure 4(a) with the spectrum number labeled (counting from left to the right). Spectrum #21 was acquired at the edge of the MoS$_2$. While the band profile remained relatively flat in the bulk of MoS$_2$ atomic layer, close to the edge both the CBM and VBM started shifting upward by ~ 0.6 ± 0.1 eV, as indicated by the transition from spectrum #19 to #21. Figure 4(c) is the color coded rendering of the real space imaging of the band profile plotted in terms of $d\left[\log\left(\frac{dI}{dV}\right)\right]/dV$. In this plot, the steepest negative slope (dark blue) occurs at the VBM while the sharp positive slope (red) occurs at the CBM; thus, their spatial traces represent the spatial variation of the CBM and VBM, respectively. The tunneling spectrum acquire at the edge (spectrum #21 in Figure 4(a)) also exhibits finite conductivity in the energy gap, reflecting the existence of metallic edge states that have been theoretically predicted.[16,18,19] Beyond the edge (starting from spectrum #22), the semi-metallic nature of graphite is also well-reflected by the tunneling spectra. The well-behaved semi-metallic features observed in the graphite region also a good indication that the tip states are smooth varying metallic states without sharp features, and the tip-to-sample distance is close enough to provide high sensitivity to the sample's density of states.

As revealed by the band profile imaging, in the bulk region, the VBM position is stably pinned at 1.83 eV below the E$_F$ (consistent with Figure 2(b)). This seems to be an intrinsic property of the planar heterojunction formed by the SL-MoS$_2$ and the underlying graphite, as all the SL-MoS$_2$ on graphite samples that we have investigated show the same property. This fact also rules out that the band bending near the edge is caused by the interface of MoS$_2$ and the graphite at the edge. Rather this is caused specifically by the metallic edge states which pins the Fermi level at the edge, just like that how the Fermi level of a 3D semiconductor would be pinned by the metallic surface states. If the edge state is symmetrically distributed, then the natural pinning position would be at the mid gap. Certainly the actual DOS of the edge states would not be so ideal but one should still expect that the pinning position to be somewhere near the mid gap. Here, the Fermi level is pinned at around 1.2 eV above VBM which is indeed close to the mid gap position of 1.07 eV above the VBM based on the quasiparticle gap of 2.15 eV. The ultra-short depletion length of 5 nm is also consistent with a heavily n-type doping.

In conclusion, by combining STM/S to measure the quasiparticle band gap and the photoluminescence measurement of the optical gap, we determine an exciton binding energy of $0.22 \pm 0.1$ eV in SL-MoS$_2$ which is smaller than the theoretical predictions. Nevertheless, we cannot rule out the possibility that narrowing of quasiparticle band gap occurs due to the formation of the impurity band. In the bulk region (i.e. interior region away from the edge) all SL-MoS$_2$ triangular islands on graphite show n-type doping with the Fermi level located at 1.8 eV above the VBM. In addition, we show that the edge is indeed metallic, as predicted by theoretical calculation. Such metallic edge states pin the Fermi level at about $1.2 \pm 0.1$ eV above the VBM with a rather short depletion length of 5 nm. Our observations not only address the key issue of exciton binding energy, but also unravel detailed information of spatial band profile of SL-MoS$_2$ on graphite, with important implications in the design and applications of single layer TMDs for electronics and optoelectronics.

**Acknowledgments**

C. Z., A. J., and C-K.S. wish to acknowlege support from the Welch Foundation grant F-1672, and NSF grant DMR-0955778. C-L. H. and L-J. Li acknowledge support from Academia Sinica (IAMS and Nano program) and National Science Council Taiwan (102-2119-M-001-005-MY3). We also thank Professor Wang Yao of University of Hong Kong for useful discussions.

**Author contributions**

All authors contributed to the intellectual content of the paper. C-L. H. and L-J. Li provided the samples. C. Z. collected and analyzed the STM data, and prepared the manuscript. A. J. carried out the photoluminescence measurements. C-K. S. advised on the experiment and provided input on the data analysis. All authors provided inputs to the manuscript.

**Competing financial interests**

The authors declare no competing financial interests.

**Figures**

**Figure 1** (a) Large area AFM image of the as-grown MoS$_2$ on HOPG; (b) STM image taken after 2 hours of annealing at around 120 °C in a UHV chamber. The SL-MoS$_2$ flake is still partially covered by unknown adsorbates. U = 3.2 V, $I_T$ = 10 pA, 200×200 nm. (c) STM image of the clean surface after proper annealing (250 °C for 10 hours). The insert profile line shows the height of a triangular flake is about 0.67 nm, which is consistent with the thickness of SL-MoS$_2$. U = 3.0 V, $I_T$ = 12 pA, 800×800 nm. (d) The zoomed-in image of a triangular SL-MoS$_2$ flake, the inset figure is a close-up image of the relatively clean edge. U = 3 V, $I_T$ = 7 pA, 100×100 nm; for inset image U = 2.7 V, $I_T$ = 7 pA, 7×7 nm.

**Figure 2** (a) The typical scanning tunneling spectra taken on the bulk area of a SL-MoS$_2$ flake. The I-V spectrum is displayed in black, while the red curve shows the corresponding *dI/dV-V* spectrum in logarithmic scale. The green dashed arrows indicate the positions of the VBM and CBM, labled as E$_v$ and E$_c$, which are equal to -1.81 and 0.30 eV, respectively. Another thershold is labed as E$_{th}$ marked by blue dashed arrow. (b) The statistical distributions based on 86 individual spectra. The upper panel is for VBM, lower panel is for CBM. The solid curves are normal distristion fittings. (c) The statistical distribution of quasipartical gap measured by STS. Mean value: 2.15 eV, standard deviation: 0.06 eV.

**Figure 3** The photoluminescence measured on the same sample after the STM investigations. The sample's temperature was varied from 79 K to 299 K using an Oxford Instruments continuous flow cryostat and temperature controller. Only spectra at 79 K (in blue) and 299 K (in black) are displayed. The sample was excited with 532 nm light in a glancing angle excitation geometry, and the PL was collected with an optical microscope. The PL was analyzed using an ARC Spectra Pro-500i spectrometer and a Si CCD detector.

**Figure 4** (a) A subset of *dI/dV* spectra that were taken along the dashed blue line in (b) from bulk of SL-MoS$_2$ to graphite. The spectrum number was labeled (counting from left to the right in the path line). Spetrum #21 (marked in (b) with red cross) was taken at the edge which shows the finite conductivity in the gap region. The dashed light blue lines indicate the zero conductance baselines for each spectrum. (b) The path along which the *dI/dV* spectra were taken shown by a black-white line. The total length is about 90 nm with a step size of 2.5 nm. The parameters used for the lock-in amplifier are same as Figure 2(a). STM image was taken with U = 2.7 V, $I_T$ = 15 pA. (c) The color coded rendering of the real space imaging of the band profile plotted in terms of $d\left[\log\left(\frac{dI}{dV}\right)\right]/dV$ where the left side is attributed to the SL-MoS$_2$. The spatial traces of the VBM and CBM are highlighted by the red and black dashed lines, respectively.

**Figure 1**

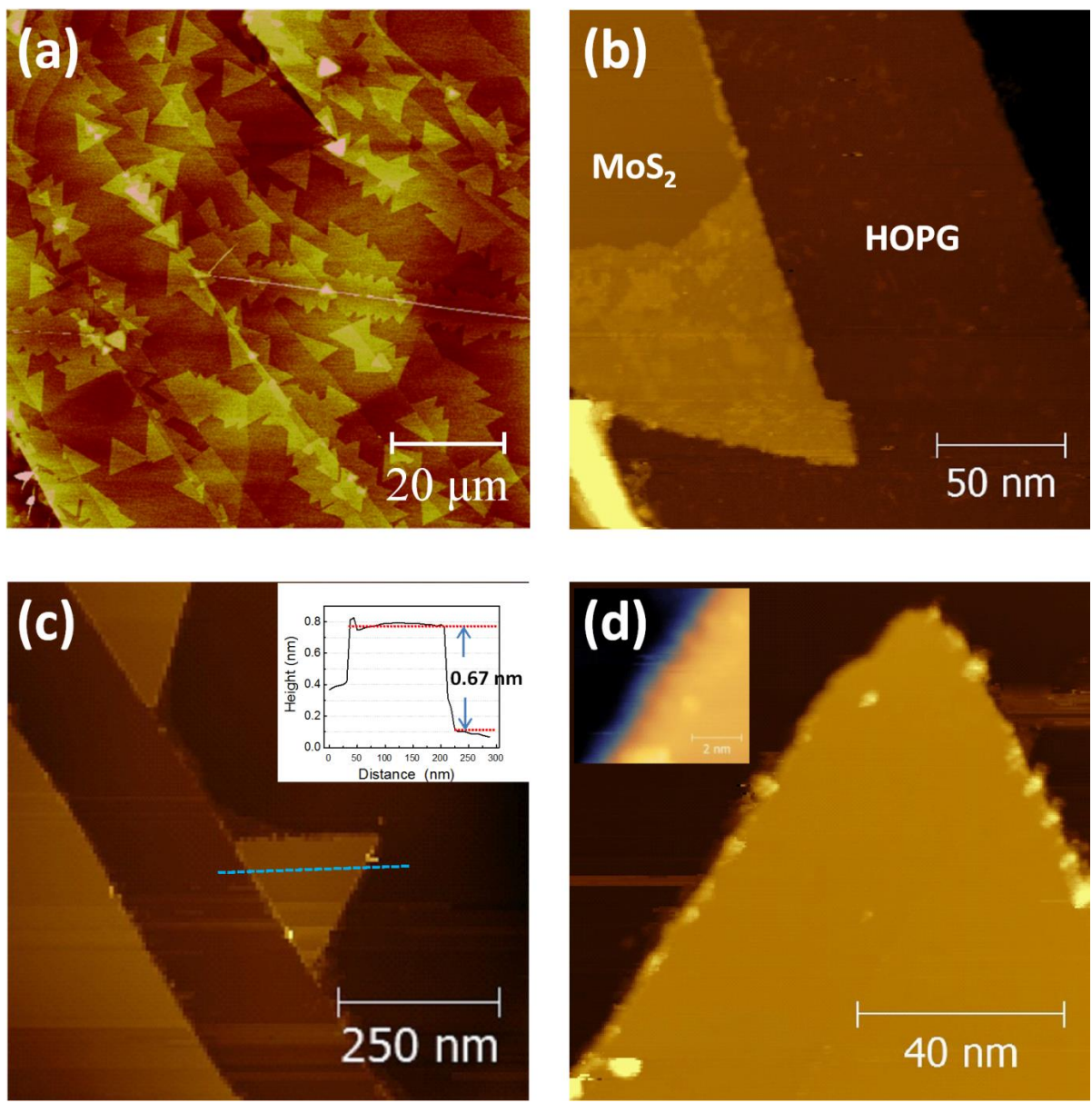

**Figure 2**

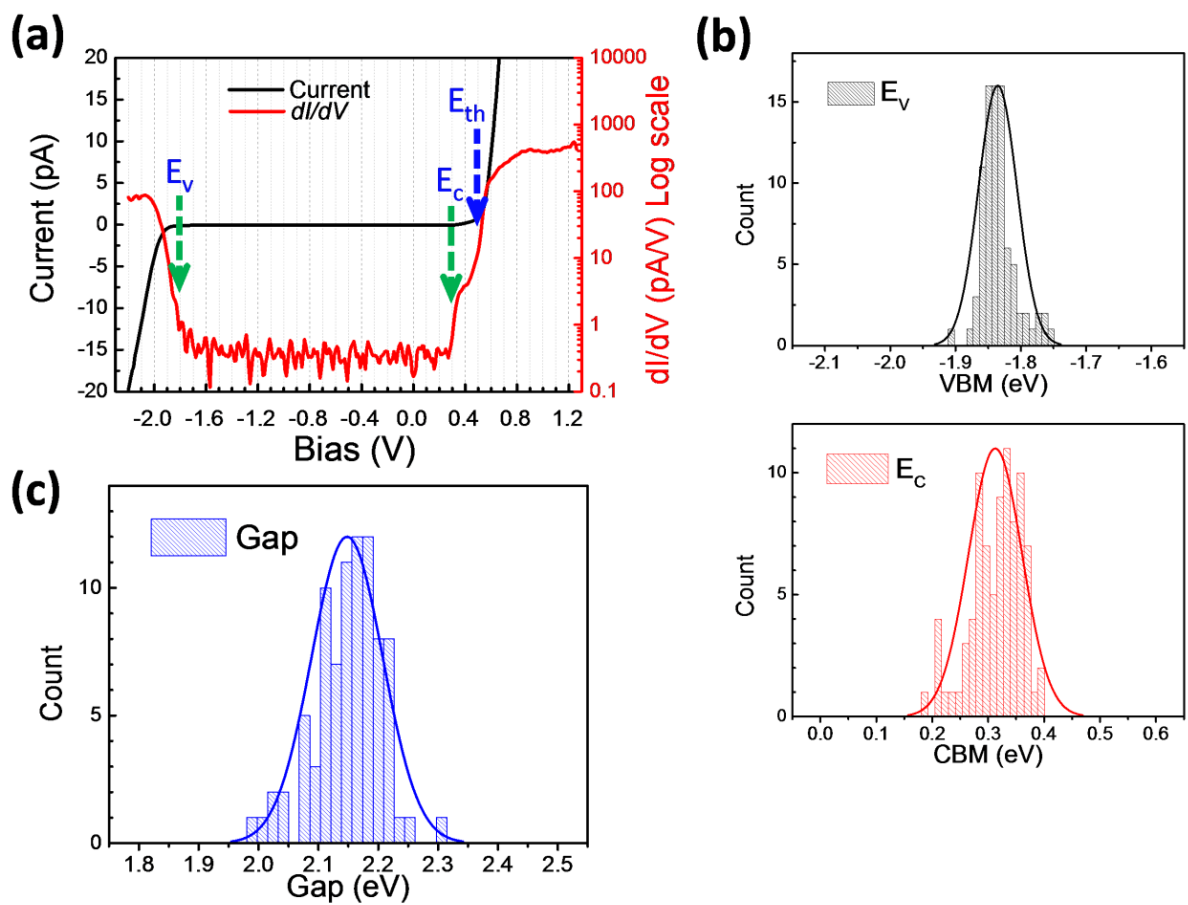

**Figure 3**

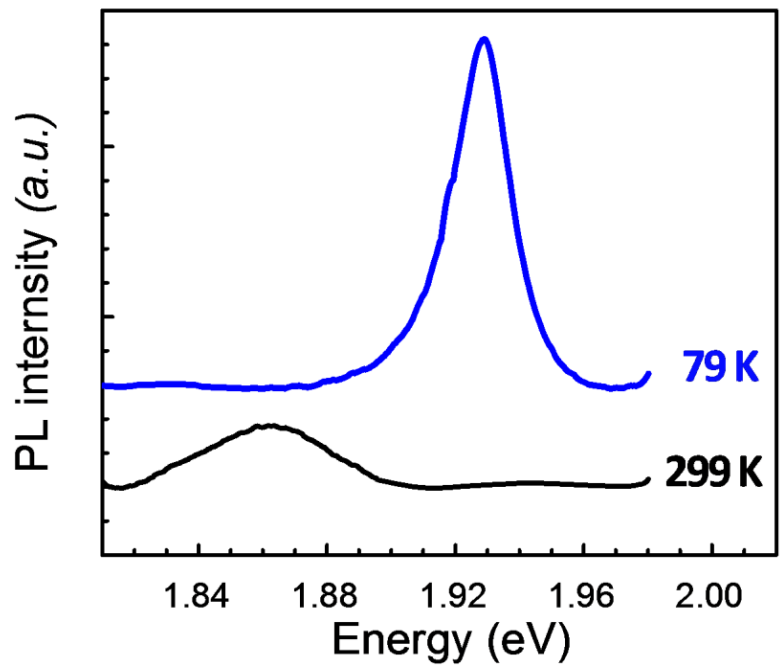

**Figure 4**

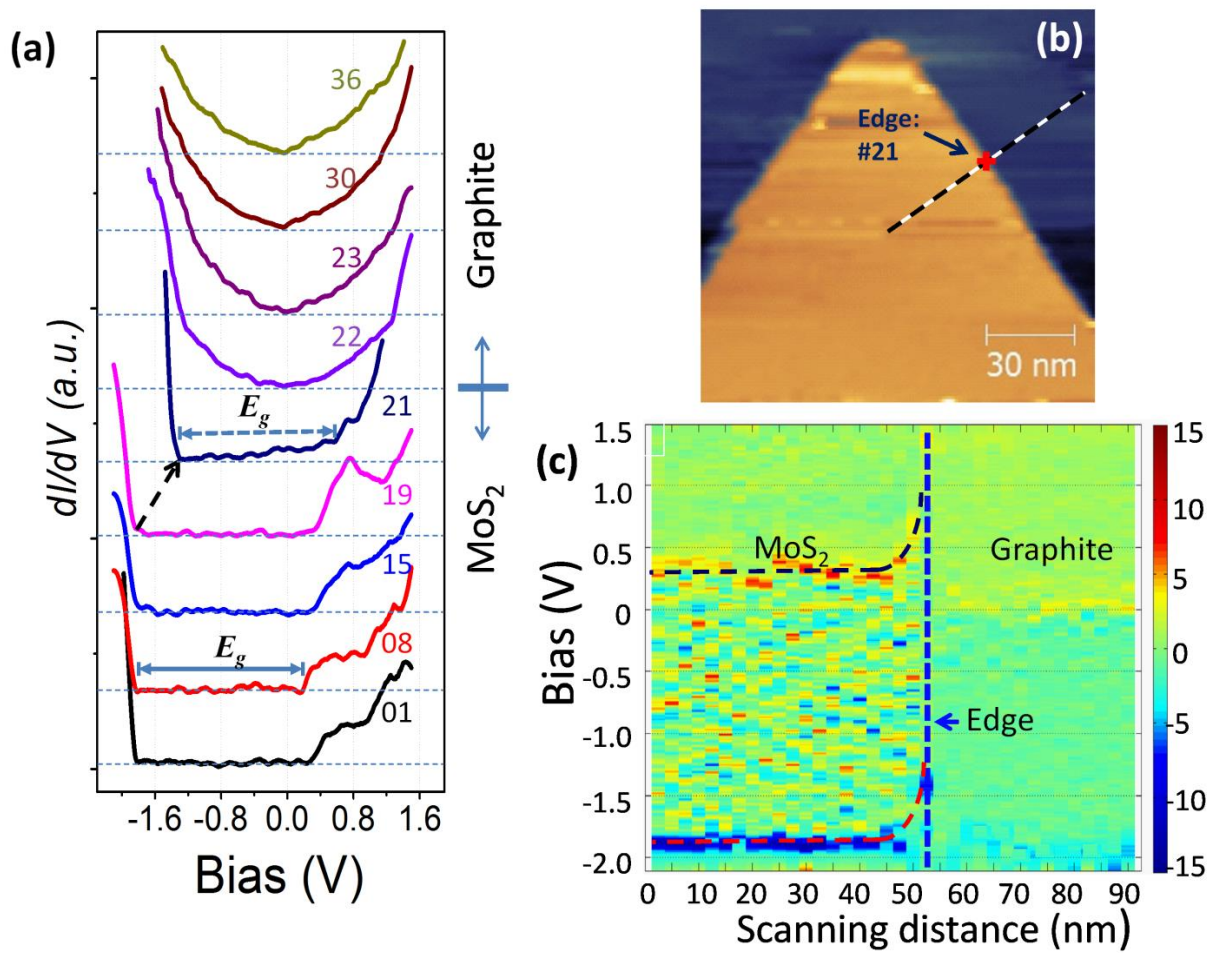